# Towards a Lattice Calculation of the Nucleon Structure Functions*

M. Göckeler[a], R. Horsley[a,b], M. Ilgenfritz[c], H. Perlt[d], P. Rakow[e], G. Schierholz[b,a] and A. Schiller[d]

[a]Höchstleistungsrechenzentrum HLRZ, c/o Forschungszentrum Jülich, D-52425 Jülich, Germany

[b]Deutsches Elektronen-Synchrotron DESY, Notkestraße 85, D-22603 Hamburg, Germany

[c]Institut für Physik, Humboldt-Universität, D-10099 Berlin, Germany

[d]Fak. f. Physik und Geowiss., Universität Leipzig, Augustusplatz 10–11, D-04109 Leipzig, Germany

[e]Institut für Theoretische Physik, Freie Universität Berlin, Arnimallee 14, D-14195 Berlin, Germany

We have initiated a programme to compute the lower moments of the unpolarised and polarised deep inelastic structure functions of the nucleon in the quenched approximation. We review our progress to date.

## 1. INTRODUCTION

Lepton–nucleon deep inelastic scattering (DIS) can be described in terms of 4 structure functions – $F_1$, $F_2$, $g_1$, $g_2$. Unpolarised beams lead to $F_1$, $F_2$ which have been studied for many years. Polarised beams are needed for $g_1$ and $g_2$. $g_1$ has only recently been measured by the EMC collaboration [1], while experiments for $g_2$ are planned e.g. at HERMES. From the Wilson operator product expansion, moments of the structure functions are related to certain nucleon matrix elements (see e.g. [2]). Thus calculating these matrix elements leads to a knowledge of the structure functions. For the unpolarised structure functions we have

$$\begin{aligned} 2\int_0^1 & dx x^{n-1} F_1(x,Q^2) \\ &= \sum_f Q^{(f)2} v_n^{(f)} + O(1/Q^2), \\ \int_0^1 & dx x^{n-2} F_2(x,Q^2) \\ &= \sum_f Q^{(f)2} v_n^{(f)} + O(1/Q^2) \end{aligned} \quad (1)$$

(with even $n$ starting at 2),where the expansion is in powers of $1/Q^2$ and

$$\langle \vec{p},\vec{s}|\mathcal{O}^{\{\mu_1\cdots\mu_n\}}|\vec{p},\vec{s}\rangle = 2v_n[p^{\mu_1}\cdots p^{\mu_n} - \mathrm{Tr}] \quad (2)$$

with

$$\mathcal{O}^{\mu_1\cdots\mu_n} = \left(\frac{i}{2}\right)^{n-1} \bar{\psi}\gamma^{\mu_1} D^{\mu_2}\cdots D^{\mu_n}\psi - \mathrm{Tr}. \quad (3)$$

Here $\psi = u$ or $d$ (quark flavour $f$, charge $Q^{(f)}$). The additional Wilson coefficient has been set equal to 1 (i.e. leading order in perturbation theory). (The normalisation is $\langle \vec{p},\vec{s}|\vec{p}_0,\vec{s}_0\rangle = (2\pi)^3 2E_{\vec{p}}\delta(\vec{p}-\vec{p}_0)\delta_{\vec{s},\vec{s}_0}$ with $s^2 = -m^2$.) The moments, eq. (1), have a parton model interpretation, being the powers of the fraction of the nucleon momentum carried by the parton: $v_n = \langle x^{n-1}\rangle$. For the polarised structure functions, on the other hand, we have

$$\begin{aligned} 2\int_0^1 & dx x^n g_1(x,Q^2) \\ &= \frac{1}{2}\sum_f Q^{(f)2} a_n^{(f)} + O(1/Q^2), \\ 2\int_0^1 & dx x^n g_2(x,Q^2) \\ &= \frac{1}{2}\frac{n}{n+1}\sum_f Q^{(f)2}(d_n^{(f)} - a_n^{(f)}) \\ &\quad + O(1/Q^2), \end{aligned} \quad (4)$$

and

$$\begin{aligned} &\langle \vec{p},\vec{s}|\mathcal{O}_5^{\{\sigma\mu_1\cdots\mu_n\}}|\vec{p},\vec{s}\rangle \\ &= \frac{a_n}{n+1}[s^\sigma p^{\mu_1}\cdots p^{\mu_n} + \cdots], \end{aligned}$$

*Combined talks presented by H. Perlt, P. Rakow and R. Horsley at LAT94, Bielefeld, Germany.

$$\langle \vec{p}, \vec{s} | \mathcal{O}_5^{[\sigma\{\mu_1]\cdots\mu_n\}} | \vec{p}, \vec{s} \rangle$$
$$= \frac{d_n}{n+1}[(s^\sigma p^{\mu_1} - s^{\mu_1} p^\sigma) p^{\mu_2} \cdots p^{\mu_n} + \cdots] \quad (5)$$

with

$$\mathcal{O}_5^{\sigma\mu_1\cdots\mu_n} = \left(\frac{i}{2}\right)^n \bar{\psi}\gamma^\sigma\gamma_5 D^{\mu_1} \cdots D^{\mu_n} \psi - \mathrm{Tr} \quad (6)$$

(for $g_1$ starting with $n = 0$, while for $g_2$ we begin with $n = 2$). It has been argued that the $n = 0$ moment of $g_2$ should be 0 [3] but it is not possible to check this sum rule on a lattice. The lowest moment of $g_1$ is interesting because it can be related to the fraction of the spin carried by the quarks in the nucleon, while $g_2$ contains not only $a_n$ (the so-called Wandzura-Wilczek contribution to $g_2$ [4]), but also $d_n$ – a twist-3 contribution.

## 2. THE LATTICE CALCULATION

### The lattice geometry

We perform our quenched QCD calculations for Wilson fermions on a $N_x N_y N_z N_t = 16^3 \times 32$ lattice. For the gauge update we used a cycle consisting of a single 3-hit Metropolis sweep followed by 16 over-relaxation sweeps using the $SU(3)$ algorithm as suggested in [5]. We repeated this cycle 20 times in order to generate a new configuration. The resulting configurations are highly independent – we see no correlations between hadronic quantities calculated on different gauge configurations.

Our calculations are carried out on the Quadrics (or APE) Q16 machine, a parallel computer consisting of 128 processors arranged in a $2 \times 2 \times 32$ periodic array. Since our lattice is $16^3 \times 32$ each node stores a lattice region $16 \times 8 \times 8 \times 1$. The fact that the block which each processor handles is only 1 lattice unit thick in the time direction means that considerable care is needed in programming the gauge update and fermion inversion. The Quadrics computer is a SIMD (Single Instruction, Multiple Data) machine which means that every processor must perform the same actions on its piece of the lattice as all the other processors. If the lattice were divided up in the simplest possible way, as illustrated in Fig. 1, then there would be a conflict when

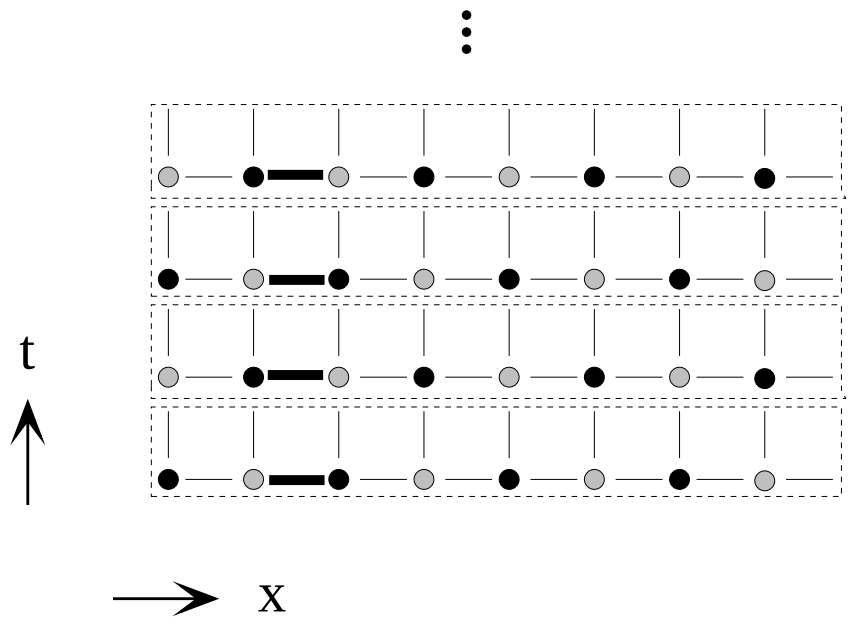

Figure 1. *A t (vertical) – x (horizontal) section of the 4-dimensional lattice fitted onto the machine geometry. Each row of points represents one processor corresponding to one t value. Links are shown as lines. Every processor works simultaneously on one link (e.g. the bold line). Even/odd lattice points are denoted by filled/shaded points.*

updating space-like links, since different processors would be making simultaneous updates within the same plaquette, which would lead to false results. Our solution to this problem is illustrated in Fig. 2 in which we have used a new *slanted* coordinate system for dividing the lattice between the different processors. Thus we set

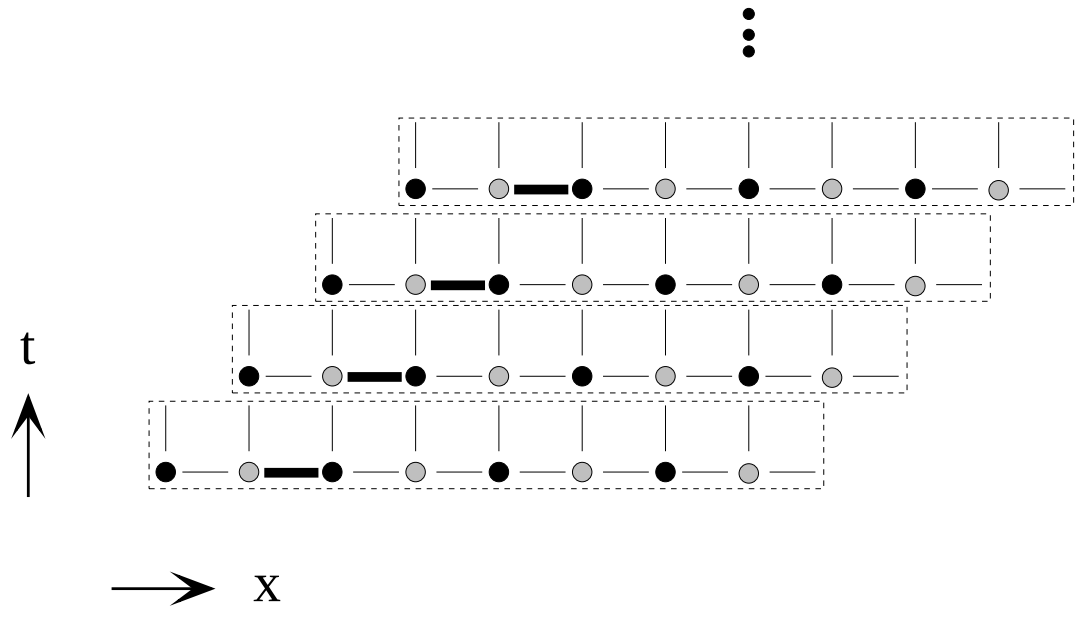

Figure 2. *As for the previous Figure, but using slanted coordinates.*

$$\begin{aligned} s_x &= (x+t) \bmod 16 \\ s_y &= y \\ s_z &= z \\ s_t &= t \end{aligned} \quad (7)$$

and use in the programme always the slanted coordinates $(s_x, s_y, s_z, s_t)$. Now there are no con-flicts during the gauge update. An additional advantage of the slanted coordinate system is that it allows algorithms that involve decomposing the lattice into even and odd sub-lattices. (Using the coordinate system of Fig. 1 this would not work – half the processors would be working on even sites and half on odd.) The even-odd decomposition is important when we invert the fermion matrix to find hadronic propagators, without this decomposition the inversion would be slower.

**Correlation functions**

The method to calculate matrix elements as in eqs. (2), (5) is well known, see [6] (we follow [7]). We first compute 2-point and 3-point correlation functions defined by

$$\begin{aligned} C_\Gamma(t;\vec{p}) &= \sum_{\alpha,\beta} \Gamma_{\beta,\alpha} \langle B_\alpha(t;\vec{p}) \bar{B}_\beta(0;\vec{p})\rangle, \\ C_\Gamma(t,\tau;\vec{p},\mathcal{O}) & \\ &= \sum_{\alpha,\beta} \Gamma_{\beta,\alpha} \langle B_\alpha(t;\vec{p}) \mathcal{O}(\tau) \bar{B}_\beta(0;\vec{p})\rangle, \end{aligned} \tag{8}$$

where $\mathcal{O}$ is the Euclideanised version of the operators in eqs. (3), (6). The derivative operator $D = \overrightarrow{D} - \overleftarrow{D}$ is defined by

$$\begin{aligned} &\overrightarrow{D}_\mu (x,y) \\ &= \tfrac{1}{2}\left[U_\mu(x)\delta_{y,x+\hat{\mu}} - U^\dagger_\mu(x-\hat{\mu})\delta_{y,x-\hat{\mu}}\right]. \end{aligned} \tag{9}$$

Using transfer matrix methods we can show that the ratio of 3- to 2- point correlation functions can be written as

$$\begin{aligned} &R(t,\tau;\vec{p},\Gamma,\mathcal{O}) \\ &= C_\Gamma(t,\tau;\vec{p},\mathcal{O})/C_{\frac{1}{2}(1+\gamma_4)}(t;\vec{p}) \\ &= \frac{1}{2\kappa}\frac{E_{\vec{p}}}{E_{\vec{p}}+m} F(\Gamma,\mathcal{J}) \end{aligned} \tag{10}$$

with

$$\begin{aligned} F(\Gamma,\mathcal{J}) &= \frac{1}{4}\mathrm{tr}\left[\Gamma N \mathcal{J} N\right], \\ N &= \gamma_4 - i\vec{p}\cdot\vec{\gamma}/E_{\vec{p}} + m/E_{\vec{p}} \end{aligned} \tag{11}$$

and $\mathcal{J}$ defined by

$$\langle \vec{p},\vec{s}|\mathcal{O}|\vec{p},\vec{s}\rangle = \bar{u}(\vec{p},\vec{s})\mathcal{J}u(\vec{p},\vec{s}). \tag{12}$$

When calculating 3-point functions it is particularly important that the baryon operator $B$ should be carefully chosen to have a low overlap with excited baryon states, in order to make the plateau region in $\tau$ as broad as possible. As our basic baryon operator we use the '$C\gamma_5$' wave-function (with $C = \gamma_4\gamma_2$ in our representation)

$$\begin{aligned} &B_\alpha(t;\vec{p}) \\ &= \sum_{\vec{x},abc} e^{-i\vec{p}.\vec{x}} \epsilon_{abc} u^a_\alpha(x)(u^b(x)C\gamma_5 d^c(x)) \end{aligned} \tag{13}$$

with two important improvements. First we use a variation of 'Wuppertal smearing' [8] namely 'Jacobi smearing' [9] in order to have an extended proton operator. Thus each quark operator in eq. (8) is replaced by

$$\psi \to \psi^S \equiv \sum_{n=0}^{N_s} (\kappa_s \overrightarrow{D})^n \psi. \tag{14}$$

We found a suitable value of $(N_s,\kappa_s)$ to be $(50, 0.21)$ which for our present $(\beta,\kappa)$ value of $(6.0, 0.155)$ gave a rms radius of about 4, corresponding roughly to $\sim 0.5$fm – half the nucleon radius. Secondly we replace each spinor by:

$$\begin{aligned} \psi &\to \psi^{NR} \equiv \tfrac{1}{2}(1+\gamma_4)\psi, \\ \bar{\psi} &\to \bar{\psi}^{NR} \equiv \bar{\psi}\tfrac{1}{2}(1+\gamma_4). \end{aligned} \tag{15}$$

We call the resulting projected wave-function the 'non-relativistic' wave function [10]. This replacement leaves quantum numbers unchanged but we would expect it to improve overlap with those baryons which have slow-moving valence quarks. Practically this means that for each baryon Green's function we invert on a smeared local source and consider only the first two Dirac components (i.e. the $2\times 2$ sub-matrix of the upper left hand corner of the full Green's function for our representation of the $\gamma$-matrices). So we only have $2\times 3$ inversions to perform rather than the usual $4\times 3$ inversions. This saves considerable computer time in the (minimal residue) inversion. In Fig. 3 we compare several baryon propagators. In each step we see an improved overlap with the proton. The projection operator is particularly effective at reducing the amplitude of the unwanted backward propagating state. Plots of the effective mass $\ln(C(t)/C(t+1))$, Fig. 4, show good plateaus for the two smeared operators with the expected proton mass 0.66(1) [11] (the wall source result given there would seem to be a little low). In particular we see that after a distance of about 4 time units we have very little trace of an excited state. In Fig. 5 we show the nucleon energy (0.77(2)) with momentum $\vec{p} \equiv \vec{p}_1 = (2\pi/16, 0, 0)$. We see that the continuum energy dispersion relation $E_{\vec{p}}^2 = \vec{p}^2 + m^2$ is obeyed.

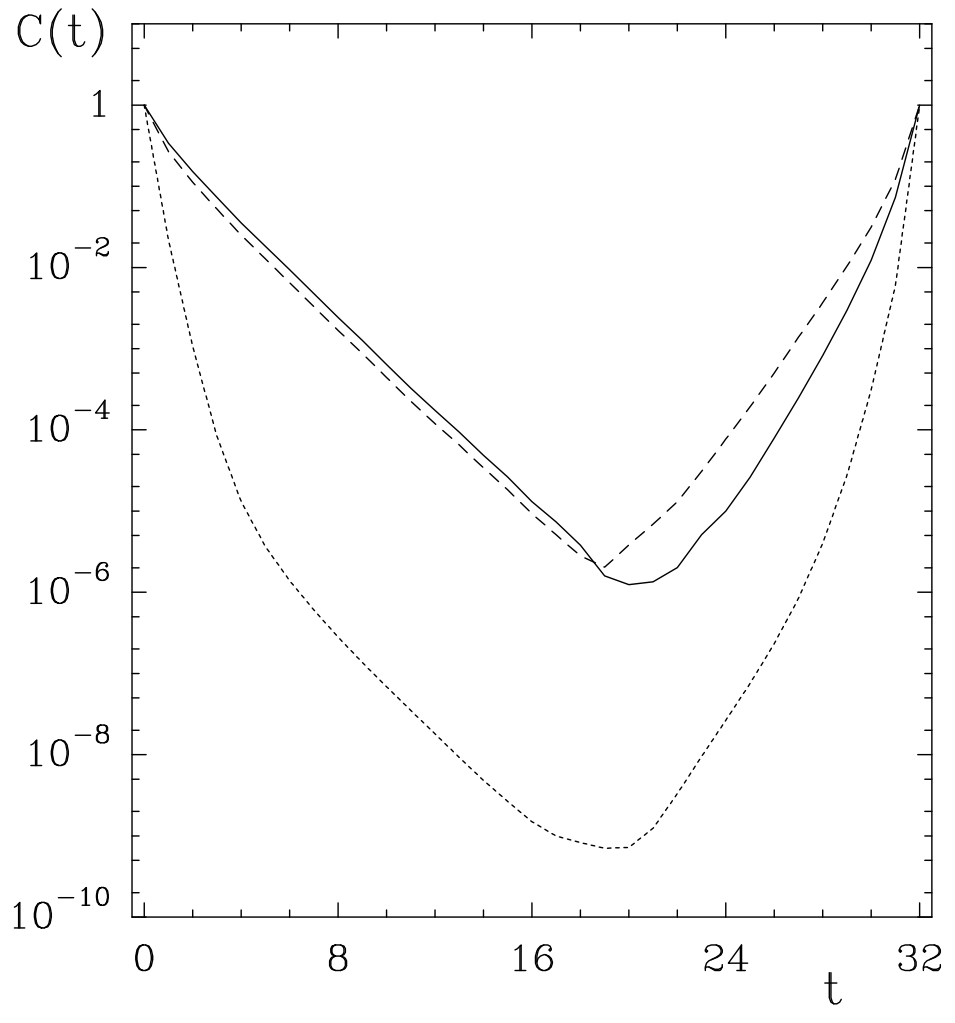

Figure 3. *Comparison between point $C\gamma_5$ wave function (dotted line, using a smaller sample size of* 100*), smeared wave function (dashed line) and smeared, non-relativistic wave function (full line) baryon correlation functions. For clarity we have dropped the (small) error bars.*

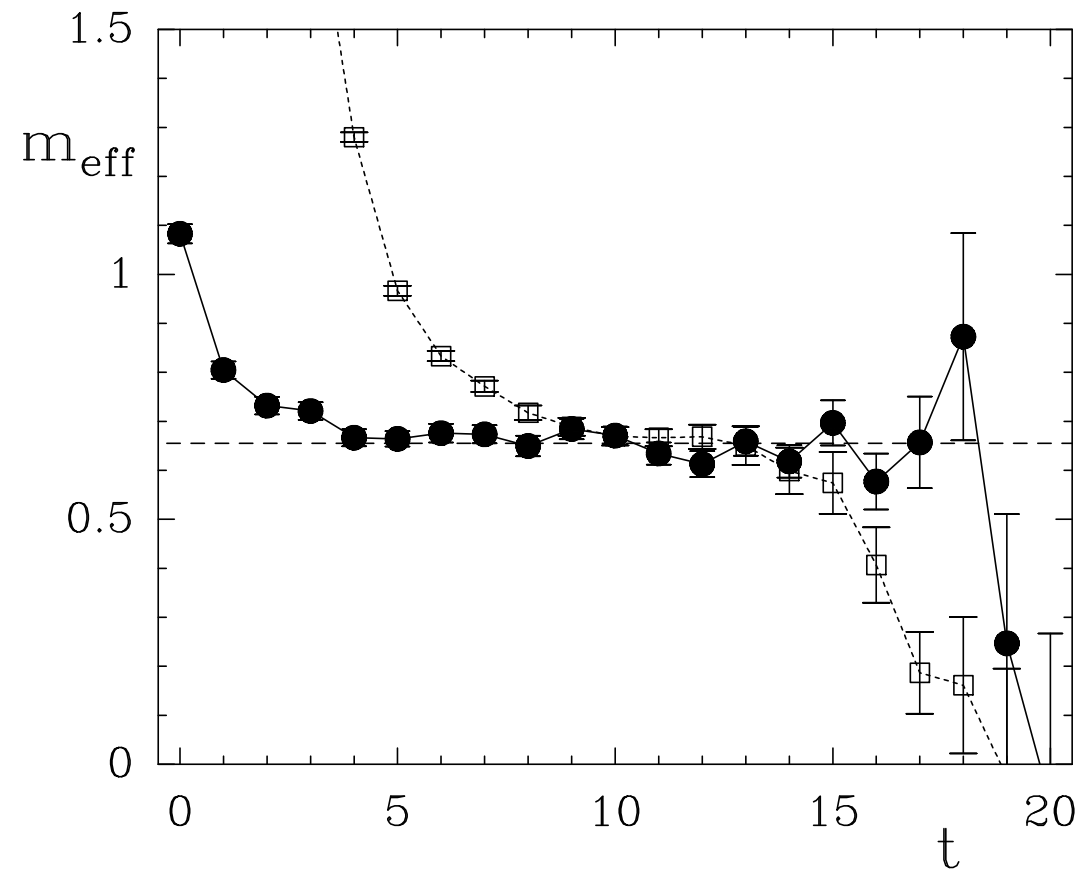

Figure 4. *Effective mass for the point $C\gamma_5$ wave function (empty square symbols) and the smeared non-relativistic wave function (filled circles).*

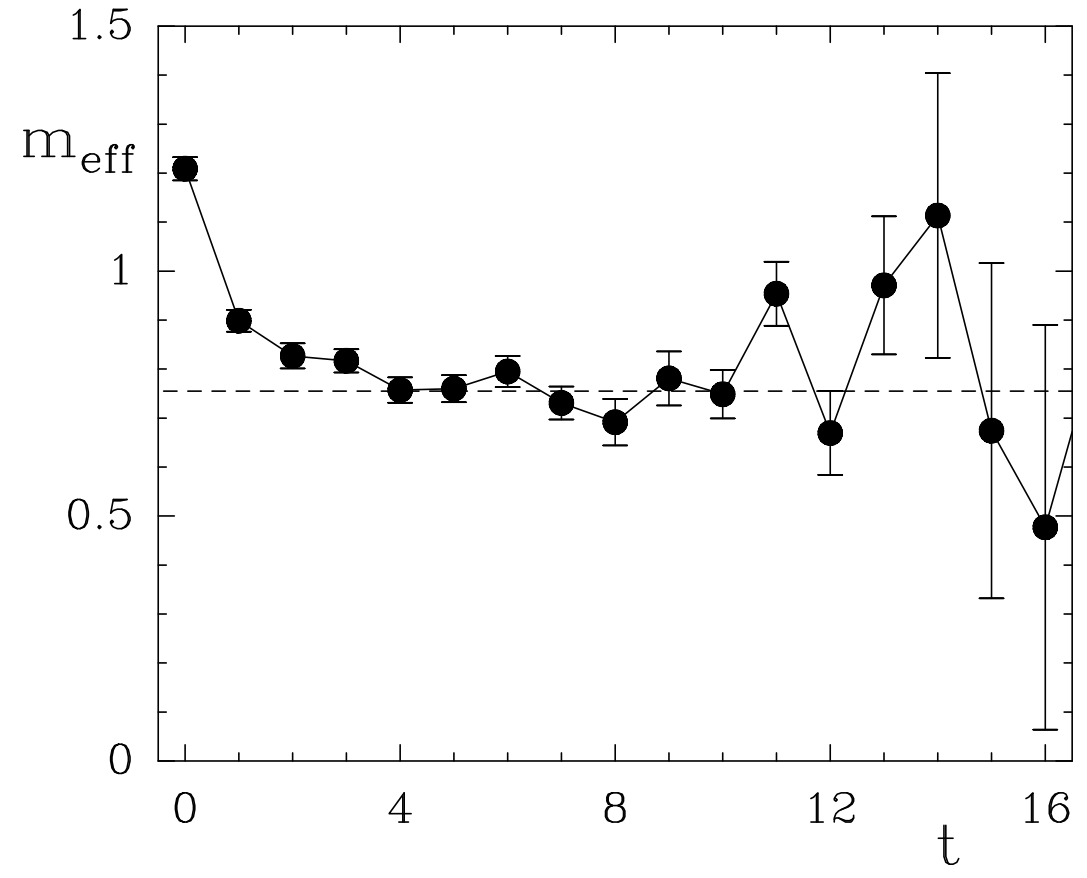

Figure 5. *Effective energy for the smeared non-relativistic wave function with momentum.*

To calculate 3-point functions we require additional Green's functions, one for each chosen $t$, $\vec{p}$ and $\Gamma$. We have fixed $t$ at 13, set $\vec{p} = \vec{0}$, $\vec{p}_1$ and have chosen two $\Gamma$ matrices, $\Gamma = \frac{1}{2}(1+\gamma_4)$, corresponding to the unpolarised case, and $\Gamma = \frac{1}{2}(1+\gamma_4)i\gamma_5\gamma_2$, corresponding to polarisation (+ - −) in the $y$ direction. We have also considered $\psi = u$, $d$ separately. This means that we must find $2 \times 2 \times 2 = 8$ (half) Green's functions. We should emphasise here that the inserted operators $\mathcal{O}$ are neither smeared nor is a non-relativistic projection used (which means that the Green's functions running to $\mathcal{O}$ are smeared/projected on one Dirac index only). The choice $t = 13$ is sufficient, larger values of $t$ lead to unacceptably large errors in the signal for $R$ – particularly for non-zero momentum. For example, test runs for $t = 17$ turned out to have $\sim O(2)$ larger errors, which roughly corresponded to the increase in the noise in the baryon correlation function from $t = 13$ to $t = 17$.

We have computed all operators up to and including 3 derivatives. At present we have generated 350 configurations. We hope for a much higher statistic and so the following results should be regarded as preliminary.

**Mixing, renormalisation**

We must now choose favourable combinations of $\sigma$, $\mu$ indices for the operators in eqs. (3), (6). Simplicity dictates that we choose indices where there is no mixing between the operator and lower dimensional operators. On the lattice this is a more severe problem than in the continuum. We have explicitly checked [12] by seeing how the operator transforms on the lattice, whether mixing occurs or not. For low derivative operators there is no problem while for the higher derivative operators mixing is excluded for certain off-diagonal combinations of the indices. However this then requires the use of non-zero momentum [13]. Note also that for the quenched theory there is no mixing with gluon operators [14].

The bare lattice operators are related to finite operators renormalised at the scale $\mu$ by

$$\mathcal{O}(\mu) = Z_{\mathcal{O}}(a\mu, g(a))\mathcal{O}(a). \qquad (16)$$

We define

$$\langle q(p)|\mathcal{O}(\mu)|q(p)\rangle = \langle q(p)|\mathcal{O}(a)|q(p)\rangle\,|^{tree}_{p^2=\mu^2}, \qquad (17)$$

where $|q(p)\rangle$ denotes a quark state of momentum $p$, and similarly for the gluonic operators. In the limit $a \to 0$ this definition amounts to the continuum, momentum subtraction renormalisation scheme. The structure functions themselves do not depend on $\mu$: the $\mu$-dependence of the renormalisation constants is compensated by the $\mu$-dependence of the Wilson coefficients. If the Wilson coefficients $c(Q^2/\mu^2, g(\mu))$ and the structure functions are evaluated at the momentum scale

$$Q^2 = a^{-2} = \mu^2 \qquad (18)$$

($\approx 2\text{GeV}^2$ here) the $a$- and $\mu$-dependence is eliminated at least on the perturbative level. In the following we will assume (18), and we will denote the 'finite' part of the renormalisation constants by $Z_{\mathcal{O}}$ without argument.

For a few cases relevant to our investigation the renormalisation constants have been computed in perturbation theory. What is known is the axial current $a_0$, where we shall take $Z_{a_0} = 0.87$ [15] and $\langle x\rangle$, $Z_{\langle x\rangle_a} = 1.0127$ [16]. We are in the process of calculating all $Z_{\mathcal{O}}$, see sect. 4. For the present in all other cases we take $Z_{\mathcal{O}} \equiv 1$. (We expect that for 0, 1, $> 1$ derivatives $Z_{\mathcal{O}} < 1, \approx 1, > 1$ respectively [13].)

## 3. RESULTS

We first consider the unpolarised structure functions. As noted previously, moments are directly related to $\langle x^n\rangle$. For the lowest moment, $\langle x\rangle$, we have no mixing problems, whether we choose off-diagonal components, $\mathcal{O}_{\{14\}}$, method $(a)$, or diagonal components, $\mathcal{O}_{44} - (\mathcal{O}_{11} + \mathcal{O}_{22} + \mathcal{O}_{33})/3$, method $(b)$, which provides another useful consistency check (and perhaps gives indications about possible lattice artifacts). For $(a)$ we have from eq. (10), $R_a = i/Z_{\langle x\rangle_a} \cdot 1/2\kappa \cdot p_1 \cdot \langle x\rangle_a$. In Fig. 6 we show $R_a$ for the $u$ quark. From eq. (10)

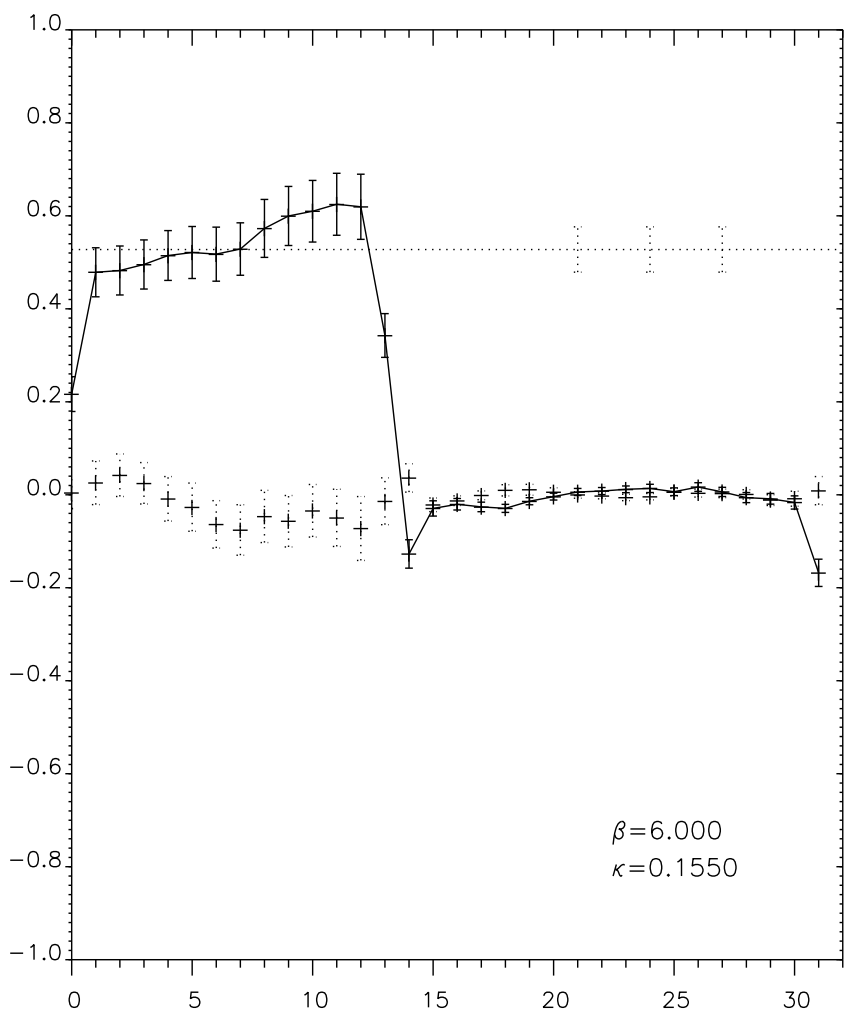

Figure 6. *The ratio of the 3-point connected correlation function to the 2-point correlation function, $R_a^{(u)}$, versus $\tau$ for $\langle x\rangle_a^{(u)}$. A linear fit, dotted line, in $4 \le \tau \le 9$ gives* $0.53(5)$. *For $R_a^{(d)}$ we have* $0.23(2)$.

we hope that in the region $0 \ll \tau \ll t = 13$ we see a plateau. In addition the dotted line is the 'conjugate signal' (here the real part of the correlation function). It should be zero, and we monitor it to give a further indication of how good our signal is. We would claim that we see a reasonable signal. A linear fit gives

$$\langle x\rangle_a^{(u)} \approx 0.42(4), \quad \langle x\rangle_a^{(d)} \approx 0.18(2). \qquad (19)$$

Similarly for $(b)$ we have $R = -1/Z_{\langle x\rangle_b} \cdot 1/2\kappa \cdot m \cdot \langle x\rangle$ shown in Fig. 7, with results

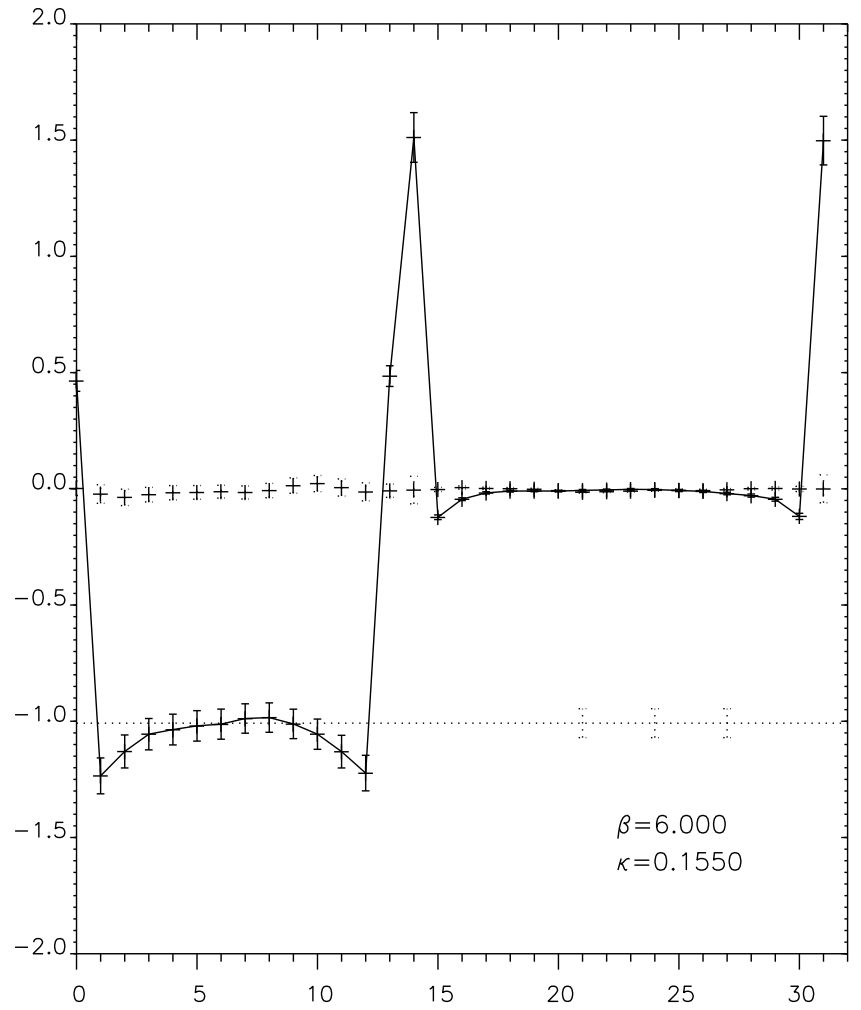

Figure 7. *$R_b^{(u)}$ versus $\tau$ for $\langle x\rangle_b^{(u)}$. A linear fit gives $-0.99(3)$. For $R_b^{(d)}$ we have $-0.45(3)$.*

$$\langle x\rangle_b^{(u)} \approx 0.47(2), \quad \langle x\rangle_b^{(d)} \approx 0.21(2). \qquad (20)$$

The methods seem to be consistent. The numbers can be compared to phenomenological parameterisations of the experimental data, e.g. from [17] ($D_-$), of 0.28, 0.11 at $Q^2 = 4\mathrm{GeV}^2$ for $u$, $d$ respectively. Analytically continuing in $n$, for $\langle x^2\rangle$ we have with $\mathcal{O}_{\{114\}} - (\mathcal{O}_{\{224\}} + \mathcal{O}_{\{334\}})/2$, $R = -1/Z_{\langle x^2\rangle} \cdot 1/2\kappa \cdot p_1^2 \cdot \langle x^2\rangle$ shown in Fig. 8, which gives

$$\langle x^2\rangle^{(u)} = 0.12(2), \quad \langle x^2\rangle^{(d)} = 0.050(8) \qquad (21)$$

to be compared with 0.08, 0.03 respectively. Both the $\langle x\rangle$, $\langle x^2\rangle$ numerical results seem to be about 50% larger than the phenomenological values as already found for larger quark masses in [6]. Finally we have computed $\langle x^3\rangle$. With $\mathcal{O}_{\{1144\}} + \mathcal{O}_{\{2233\}} - \mathcal{O}_{\{1133\}} - \mathcal{O}_{\{2244\}}$, $R = 1/Z_{\langle x^3\rangle} \cdot 1/2\kappa \cdot E_{\vec{p}_1} p_1^2 \cdot \langle x^3\rangle$ is shown in Fig. 9. Although there are large fluctuations in the data, the values

$$\langle x^3\rangle^{(u)} = 0.031(8), \quad \langle x^3\rangle^{(d)} = 0.0076(47) \qquad (22)$$

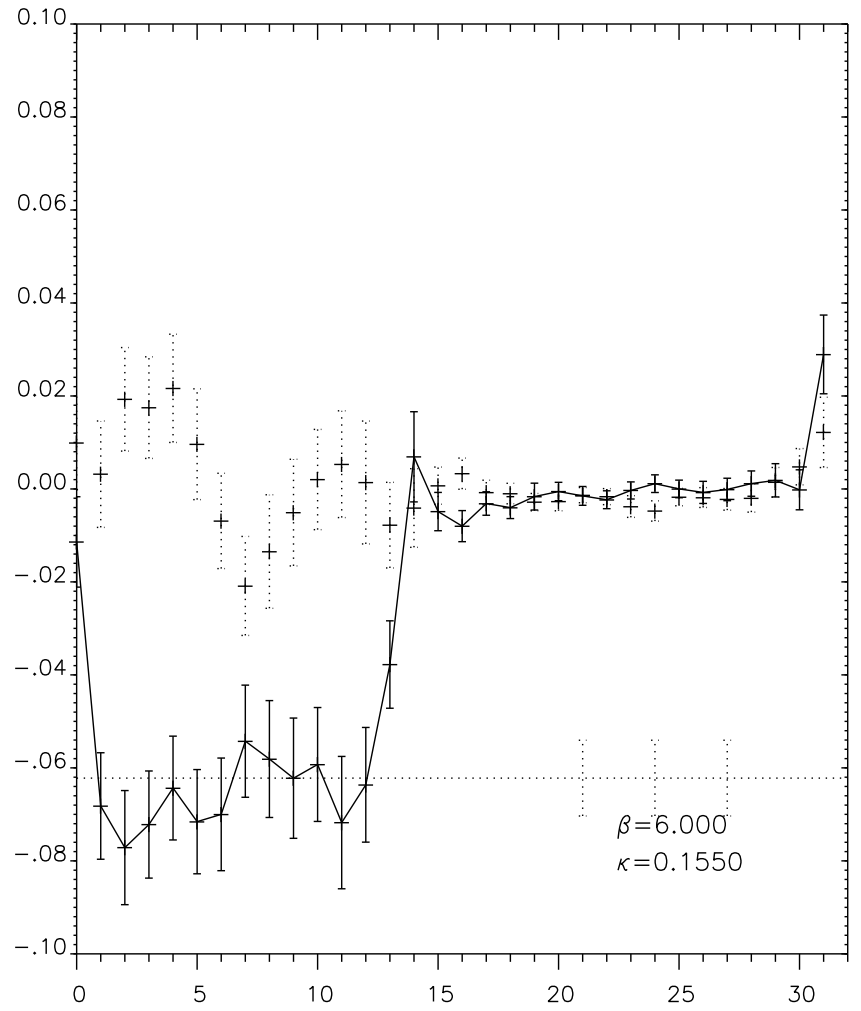

Figure 8. *$R^{(u)}$ versus $\tau$ for $\langle x^2\rangle^{(u)}$. A linear fit gives $-0.062(8)$. For $R^{(d)}$ we have $-0.025(4)$.*

are consistent with the phenomenological results of 0.03, 0.009.

We now turn to the polarised cases and start with $a_0$, which is the matrix element of the axial current. In Fig. 10 we show $R^{(u)}$ for $\mathcal{O}_{5;2}$ which gives $R = i/Z_{a_0} \cdot 1/2\kappa \cdot m/2E_{\vec{p}} \cdot a_0$ with $\vec{p} = \vec{0}$. Conventionally we set $a_0 = 2\Delta q$. A linear fit gives

$$\Delta q^{(u)} \approx 0.86(5), \quad \Delta q^{(d)} \approx -0.27(2). \qquad (23)$$

These numbers are to be compared with the experimental results, e.g. as given in [18] of 0.80(4), −0.46(4) respectively and −0.13(4) for the strange quark. For the $d$ quark, at least, the result seems a little large. Non-singlet quantities are likely to be more independent of sea quark contributions (which are absent in the quenched approximation). For the axial decay constant (from $\beta$ decay) we have experimentally $\Delta u - \Delta d = g_A \approx 1.26$. Our result $\approx 1.13$ seems a little small compared to this.

For comparison we show in Fig. 11 a calculation of the same quantity, but with non-zero momentum. We see a degradation of the signal by a factor of roughly $O(2)$. (To obtain a comparable signal as for the zero momentum case would thus require $O(4)$ times more configurations.) Nevertheless a consistent result with $\vec{p} = \vec{0}$ is obtained. We find 0.88(12), −0.31(6) for $u$, $d$ respectively.

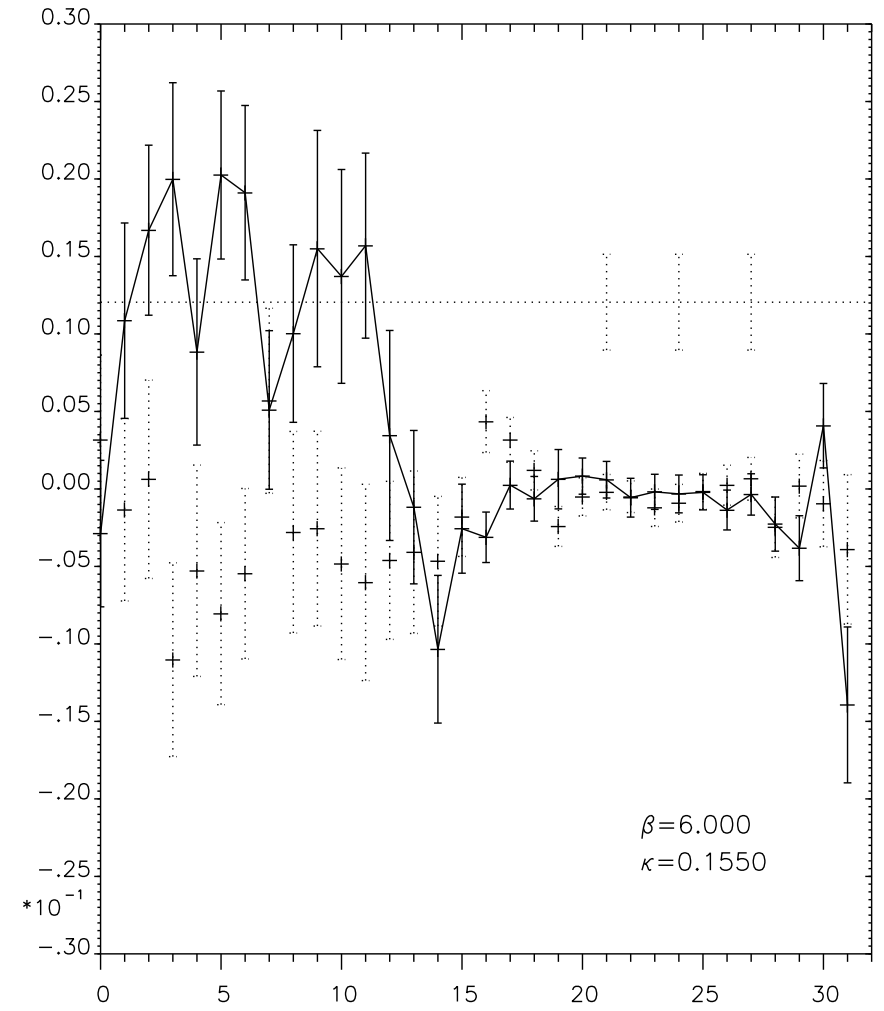

Figure 9. *$R^{(u)}$ versus $\tau$ for $\langle x^3\rangle^{(u)}$. A linear fit gives* $0.012(3)$*. For $R^{(d)}$ we have* $0.0029(18)$.

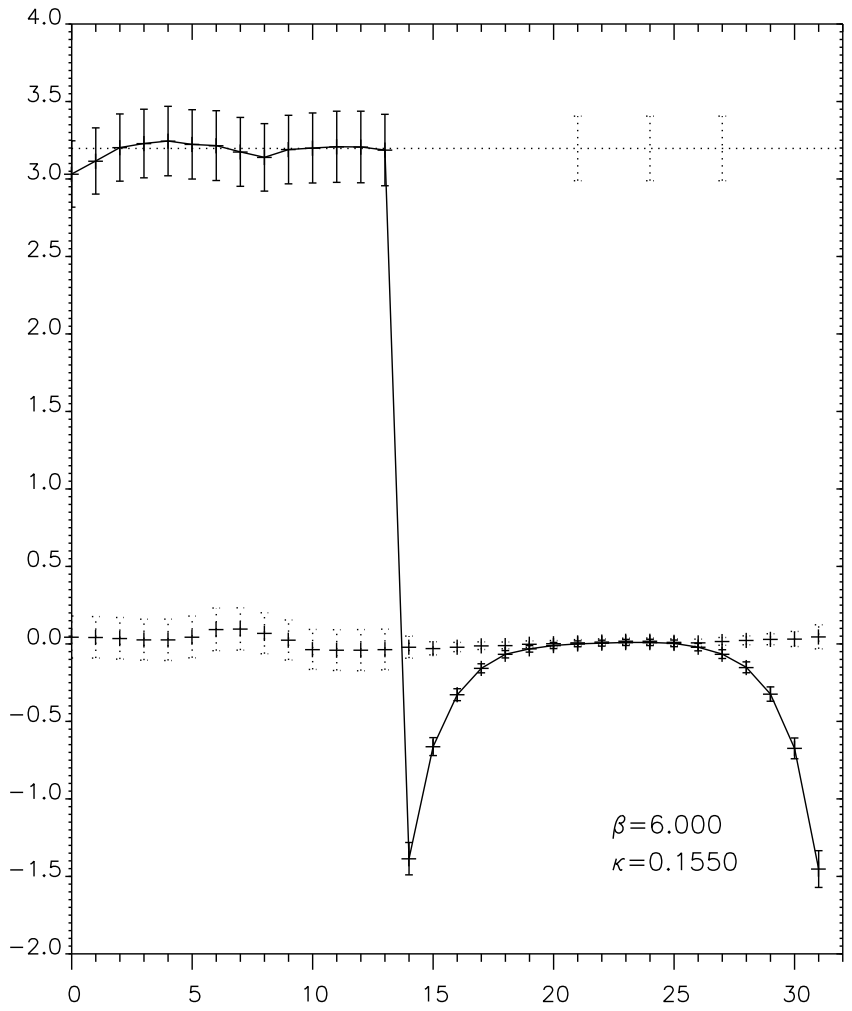

Figure 10. *$R^{(u)}$ versus $\tau$ for $a_0^{(u)}$. A linear fit gives* $3.2(2)$*. For $R^{(d)}$ we have* $-0.99(8)$.

Next we consider $a_2$. To avoid mixing problems we choose the operator $\mathcal{O}_{5;\{214\}}$ so the matrix element needs $\vec{p} = \vec{p_1}$. We have $R = i/Z_{a_2} \cdot 1/2\kappa \cdot 1/6 \cdot mp_1 \cdot a_2$ shown in Fig. 12. As expected due to the complexity of the operator, the signal is at present not so good. Fits lead to $a_2^{(u)} \approx 0.14(3)$, $a_2^{(d)} \approx -0.0047(12)$, which gives

$$\int_0^1 dx x^2 g_1 \approx \begin{cases} 0.016(3) & \text{(proton)} \\ 0.0034(10) & \text{(neutron)} \end{cases} \tag{24}$$

(for the neutron $u \leftrightarrow d$). This is to be compared to the EMC result of 0.022 for the proton.

For $d_2$ we choose $\mathcal{O}_{5;[2\{1]4\}}$ and show the ratio $R = 1/Z_{d_2} \cdot 1/2\kappa \cdot 1/3 \cdot mp_1 \cdot d_2$ in Fig. 13. An estimate gives $d_2 \approx -0.18(2)$ for the $u$ quark and $\approx 0.0036(11)$ for the $d$ quark. Thus from eq. (4) we have

$$\int_0^1 dx x^2 g_2 \approx \begin{cases} -0.013(2) - 0.010(2) \\ -0.003(1) - 0.002(1) \end{cases} \tag{25}$$

(proton/neutron) where the first number comes from $a_2$ (twist 2), while the second comes from $d_2$ (twist 3). They are of the same order of magnitude. Two other theoretical estimates have been made, using bag models [19] and sum rules [20]. Sum rule results suggest that for the proton $d_2$ is very small. At present we seem to have a somewhat larger result.

## 4. RENORMALISATION

Parallel to our numerical work we have begun to compute the renormalisation constants for all twist-2 operators up to spin 4, to one loop order. Though this is a well defined task, it is computationally voluminous due to complications imposed by the finite periodic volume and the hypercubic symmetry. In particular the vertices associated with the higher dimensional operators are of unprecedented complexity. To master this task we are currently developing packages of computer algebraic programmes using *Mathematica* and *Maple*.

### Gluon Operators

The operator we have considered first is the gluon operator

$$\mathcal{O}^g_{\mu_1\mu_2} = \mathrm{Tr} F_{\mu_1\alpha} F_{\mu_2\alpha}, \tag{26}$$

where

$$F_{\mu\nu} = \frac{1}{8iga^2} \sum_{\mu,\nu=\pm} (U_{\mu\nu} - U^\dagger_{\mu\nu}) \tag{27}$$

and $U_{\mu\nu}$ is the plaquette operator. This operator gives us the lowest moment of the gluon distribution. We need to expand $F_{\mu\nu}$ in terms of powers of the coupling constant. Writing [21]

$$U_{\mu\nu} = \exp[i(g\phi^{(1)}_{\mu\nu} + g^2\phi^{(2)}_{\mu\nu} + g^3\phi^{(3)}_{\mu\nu} + \cdots)], \tag{28}$$

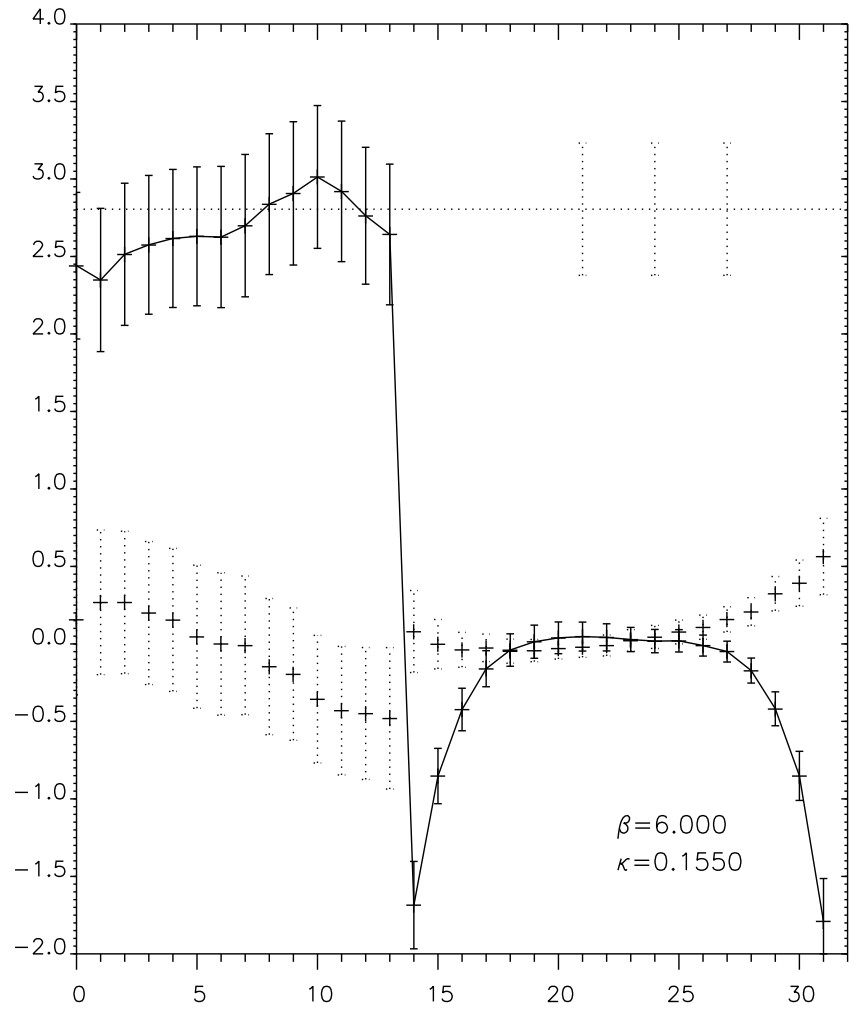

Figure 11. *$R^{(u)}$ versus $\tau$ for $a_0^{(u)}$ with momentum. A linear fit gives* 2.8(4). *For $R^{(d)}$ we have* −0.98(19).

where $\phi^{(k)}_{\mu\nu}$ is of order $k$ in the gauge potential, we obtain

$$F_{\mu\nu} = \frac{1}{4a^2}\sum_{\mu,\nu=\pm}(\phi^{(1)}_{\mu\nu} + \frac{ig}{2}(\phi^{(2)}_{\mu\nu} - \phi^{(2)\dagger}_{\mu\nu}) + \frac{g^2}{2}(2\phi^{(3)}_{\mu\nu} - \frac{1}{3}\phi^{(1)\,3}_{\mu\nu}) + \cdots \quad (29)$$

The calculation of the loop integrals is done in two parts. Following [22] we write

$$I(p)_{\mu_1\mu_2} = (I_{\mu_1\mu_2}(p) - \tilde{I}_{\mu_1\mu_2}(p)) + \tilde{I}_{\mu_1\mu_2}(p), \quad (30)$$

where

$$\tilde{I}_{\mu_1\mu_2}(p) = \sum_{n=0}^{2}\frac{p_{\alpha_1}\cdots p_{\alpha_n}}{n!}\frac{\partial^n}{\partial p_{\alpha_1}\cdots\partial p_{\alpha_n}}I_{\mu_1\mu_2}(p)|_{p=0}. \quad (31)$$

The first term in eq. (30) is ultraviolet finite and is computed in the continuum, while the second term is ultraviolet divergent and is computed on the lattice. Both terms have infrared divergences. They are regulated by dimensional regularisation and cancel out in the sum. The calculations are done in Feynman gauge. The programme has been developed to such a level that all what is needed as input is to state the Feynman rules in symbolic form, both for the continuum and the lattice part of the calculation. The lattice integrals are reduced to a few integrals that can be computed to arbitrary precision. If we write

$$Z_g = 1 - g^2 N_c B_g, \quad (32)$$

we find

$$B_g = -\frac{3}{64} - \frac{5}{48\pi^2} - \frac{5}{16}Z_0 - \frac{1}{16}Z_1 + \frac{1}{8N_c^2}, \quad (33)$$

where $Z_0 = 0.15493339$, $Z_1 = 0.10778131$. This is, as all our results, in the quenched approximation and agrees with [16]. At $\beta = 6.0$ this gives $Z_g = 1.296$. The calculation can be extended without any difficulties to higher moments of the gluon distribution function.

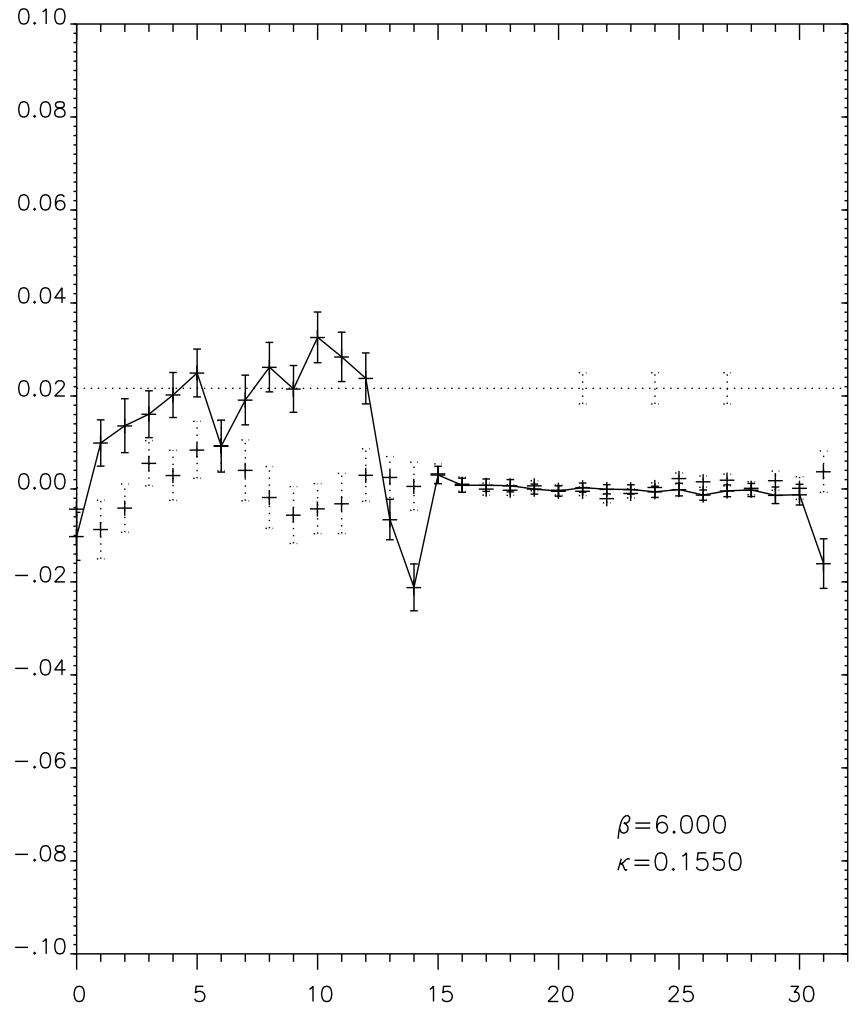

Figure 12. *$R^{(u)}$ versus $\tau$ for $a_2^{(u)}$. A linear fit gives* 0.020(4). *For $R^{(d)}$ we have* −0.0066(17).

**Quark Operators**

The calculation of the renormalisation constants for the quark operators is in progress. Here the main problem is to compute the lattice integrals over the fermion propagators to a sufficient accuracy. Like in the gluonic case we found that it is possible to reduce all integrals to just a few basic ones. We hope to report our results shortly.

## 5. DISCUSSION

We have presented here an outline of our programme for calculating low moments of polarised and unpolarised structure functions. At present we have only one $\kappa$ value, which represents rather a heavy quark mass of about 66MeV. We need at least one more $\kappa$ value (two would be better) to be able to extrapolate to a more physical quark mass. Finally, although also not reported here, we are monitoring the gluon moments (from $FF$ and $FDDF$) and calculating the hadron mass spectrum, electro-magnetic form factors (vector current), the scalar (sigma term) and pseudoscalar density (for the $g_{NN\pi}$ coupling), all at non-zero momentum transfer [12].

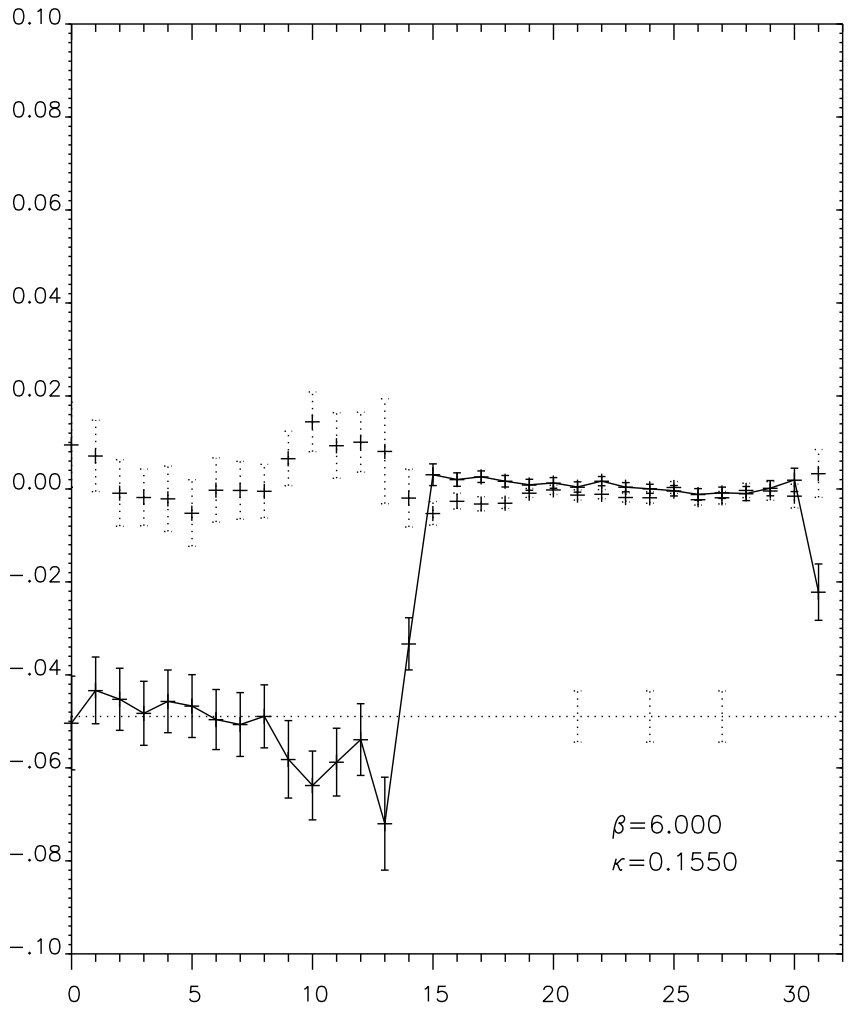

Figure 13. *$R^{(u)}$ versus $\tau$ for $d_2^{(u)}$. A linear fit gives* $-0.049(6)$. *For $R^{(d)}$ we have* $0.010(3)$.

## ACKNOWLEDGMENTS

The numerical calculations were performed on the APE at Bielefeld University and the APE at DESY (Zeuthen). We wish to thank both institutions for their support and in particular the system managers M. Plagge and H. Simma for their help.